\documentclass[12 pt,a4paper]{extarticle}
\usepackage{0style}
\begin{document}
\title{The Role of Geographic Spreaders in Infectious Pattern Formation and Front Propagation Speeds}
\author[1*,2]{Shuolin Li}
\author[3]{Craig Henriquez}
\author[1,2]{Gabriel Katul}
\affil[1]{Nicholas School of the Environment and Earth Science, Duke University, Durham, NC, USA}
\affil[2]{Department of Civil and Environmental Engineering, Duke University, Durham, NC, USA}
\affil[3]{Department of Biomedical Engineering, Duke University, Durham, NC, USA}
\affil[*] {Email: shuolin.li@duke.edu}
\maketitle
\thispagestyle{fancy}

\begin{abstract}
The pattern formation and spatial spread of infectious populations are investigated using a kernel-based Susceptible-Infectious-Recovered (SIR) model applicable across a wide range of basic reproduction numbers $R_o$. The focus is on the role of geographic spreaders defined here as a portion of the infected population ($\phi$) experiencing high mobility between identical communities. The spatial organization of the infected population and invasive front speeds ($c_{max}$) are determined when the infections are randomly initiated in space within multiple communities. For small but finite $\phi$, scaling analysis in 1-dimension and simulation results in 2-dimensions suggest that $c_{max}\sim (1-\phi) \gamma (R_o-1) \sigma$, where $\gamma$ is the inverse of the infectious duration, and $\sigma^2$ is the variance of the spatial kernel describing mobility of long-distance spreaders across communities.  Hence, $c_{max}$ is not significantly affected by the small $\phi$ though reductions in $\phi$ act as retardation factors to the attainment of $c_{max}$. The $\sigma$ determines the spatial organization of infections across communities. When $\sigma >5dr$ (long-distance mobility, where $dr$ is the minimum spatial extent defining adjacent communities), the infectious population will experience a transient but spatially coherent pattern with a wavelength that can be derived from the spreading kernel properties. 
\end{abstract}

\section*{Keywords}
Geographic Spreaders, Integro-Differential Equations, Spatial Pattern, Susceptible-Infectious-Recovered model
\clearpage

\section{Introduction}
%----------- Brief motivation
Spatial models of infectious disease transmission remain the optimal viable 'in-silico' tactic where the movement of infected individuals can be combined with a quantitative description of the infection process and disease properties to explore appropriate intervention responses \citep{riley2007large,riley2015five,grassly2008mathematical, sanz2014dynamics, valdano2015analytical, keeling2011modeling}.  The mobility of infectious individuals in a susceptible population may lead to a recognizable spatial pattern, which when understood, can serve as a basis for concentrating testing capacity or planning control strategies. For this reason, the emergence of coherent spatial patterns in epidemiology, even in the most idealized settings where the pattern is endogenous to the model system describing the spread, continue to draw attention  \citep{murray2001mathematical,eisinger2008spatial,boerlijst2010spatial,sun2016pattern, soriano2018spreading}.

%---------- What are the patterns expected
Pattern formation and deterministic models (e.g. activation-inhibition, reaction-diffusion, networking) in infectious disease have a long tradition \citep{belik2011natural, soriano2018spreading, koher2019contact}, and reviewing all their aspects is well beyond the scope of a single study.  However, these approaches delineate two broad classes of spatial patterns \citep{murray2001mathematical,fuentes2003nonlocal, sun2016pattern}. The first is stationary patterns that remain unchanged or stable over an extended period of time when referenced to the duration of the outbreak. These patterns often exhibit intermittent areas of a high density of infectious individuals. The locally concentrated areas may favor persistence of a disease and are of logical concern to epidemiology. The other, often labeled as spatio-temporal patterns, change over time either as (i) quasi-periodic or oscillatory waves \citep{sherratt2008periodic} or as (ii) spatio-temporal (intermittent) chaos \citep{fuentes2003nonlocal,sun2016pattern}.  The work here shows that spatial patterns of infections with dynamical features that share some resemblance to these two categories can emerge. These features depend on the nature of the local competitive interaction (e.g. mass action within a community) as well as on the spatial extent of the competitive range. 

%--------- What do we expect in this work
Compartment models with mass action allow infectious individuals to access all the susceptible population within an isolated community \citep{girvan2002community}.  For clarity, we label the space occupied by such a community as box so as to distinguish it from the term 'compartment' commonly used when classifying population into susceptible, infected, or recovered. The spatial domain is assumed to be much larger than the size of the box so that the box center may be identified by coordinates representing the box location.  This assumption allows space to be treated as a continuous variable. Long-distance mobility of infectious individuals from one box to another, however, leads to interaction with another susceptible population that is non-local in space, which we naively label as geographical spreaders. The transport time of infectious individuals is assumed to be much faster than recovery or removal time scale so that time may also be treated as a continuous variable.  The geographical spreaders share some similarity with the commonly used term 'superspreader' \citep{small2006super}.  A precise definition of superspreader remains elusive and often defining syndromes are used. In its broadest epidemiological sense, a superspreader has the propensity to infect a larger than an average number of susceptible individuals (where mobility is only one of several factors resulting in elevated infectious rates).  Pragmatically, a superspreader is an individual who can infect more individuals than say a number constrained by the basic reproduction number $R_o$ \citep{galvani2005dimensions}. From this pragmatic perspective, geographical spreaders and superspreaders share common attributes but geographical spreaders do not encompasses biological, behavioral, and environmental variables relevant to disease transmission. For this reason, we only focus on infectious individuals with high mobility to distinguish them from the widely used superspeader term.  This focus on spatial spread of individuals also ignores numerous other factors such as those associated with particular events where a large congregation of individuals in a small area (colloquially labeled a superspreader event) facilitate infectious spread \citep{kuebart2020infectious}.  The choice to associate high mobility only with geographical spreaders in compartment models simplifies the mathematical treatment for continuous space-time models.  It leads to well-recognized integro-differential equations (IDE) \citep{lefever1997origin,fuentes2003nonlocal,fuentes2004analytical,guo2020spatial,keeling2011modeling,murray2001mathematical} that may be viewed as a hybrid between local diffusion-based schemes \citep{lang2018analytic} and mobility on networks \citep{pastor2015epidemic}, where long-distance movement is ubiquitous.   

In the IDE approach, mobility of geographical spreaders is accommodated using a spatial kernel that describes the probability $p(x-x',y-y')$ of infectious individuals to travel from point defined by Cartesian coordinates $(x, y)$ to $(x', y')$. As such, detailed knowledge about long-range connections can, in principle, be accommodated using non-homogeneous spreading kernel functions. By “non-homogeneous”, we mean that the spreading kernel is no longer dependent on relative distance between points (x,y) and (x',y') but also on the position or origin (x,y).  This situation arises when the statistics of the spreading kernel (e.g. kernel variance) vary with position instead of relative distance $r=\sqrt{(x-x')^2+(y-y')^2}$).  Likewise, spreading kernels can be non-isotropic with certain preferential pathways breaking the symmetry of the movement. The $p(x-x',y-y')$ can be made time-dependent i.e. $p(x-x',y-y',t-t_o)$ to reflect different mobility patterns during the course of a day (daytime versus night-time), week (weekday versus weekends), or season (summer versus winter) relative to a reference time $t_o$.  The IDE approach proved effective in modeling fast propagation fronts of biological systems such as vegetation migration e.g. invasion by extremes, vegetation patterning in dryland ecosystems, the 'chic and hen' pattern in seedlings around adult trees in tropical ecosystems, among others \citep{lefever1997origin,thompson2008plant,thompson2008role,thompson2009secondary,thompson2009spatial,meron2018patterns}. The IDE approach proposed here is shown to exhibit qualitatively similar invasion thresholds reported in reaction-diffusion processes operating in meta-population models with heterogeneous connectivity patterns \citep{colizza2007invasion}.  Moreover, the approach can accommodate (indirectly) bidirectional movements between 'base' and 'destination' locations, which have been shown to generate some saturation attack velocities in mobility networks \citep{belik2011natural, soriano2018spreading}. More recently, it has been demonstrated that the IDE approach is able to reproduce measured spectral (and multi-fractal) space-time (and intermittent) properties of COVID-19 infections data for a full year across a wide range of scales (from county to continent) \citep{Genge2023321118}. However, the mathematics that govern the spread and recession speeds and the concomitant pattern formation during the outbreak have not been investigated and motivate the work here.

%---------- What is the goal of this work
The main objective then is to use the IDE approach to establish links between the properties describing the spatial kernel of geographical spreaders and (i) transient spatial pattern formation of infectious individuals during the outbreak, and (ii) invasive and recession speeds for the disease. These links are tested across a wide range of basic reproduction numbers. The specific questions to be addressed are: (i) What spatial patterns arise when multiple communities are randomly infected resulting in interfering wavefronts and (ii) what is the role of the spreading kernel properties (its variance and spatial extent) and the fraction of geographical spreaders in a community on spatial patterning of infectious individuals.  The overall impact of geographical spreaders on compartment models employing only mass action when describing encounters between infectious and susceptible populations is also analyzed and discussed.  

\section{Theory}
A brief review of standard compartment models and their extension to include spatial mobility by diffusion are covered to introduce nomenclature.  The use of IDE to expand diffusion schemes along with key mathematical properties for spatial pattern formation and front propagation are then developed for simplified kernel shapes.

\subsection{Definitions and Nomenclature}
A generic mathematical model of disease outbreak, known as the Susceptible-Infectious-Recovered (SIR) model, subdivides the total population $N_o$ residing in a closed box into 3 distinct compartments: susceptible ($S$), infectious ($I$), and recovered or removed ($R$). Infectious individuals spread the disease to a susceptible individual and remain in the infectious compartment for a given period before moving into the recovered (or removed) compartment. Individuals in the recovered or removed compartment are assumed to be immune (or removed) from the population.  The dynamical system describing the SIR equations is given as \citep{kermack1927contribution,hethcote2000mathematics,murray2007mathematical,keeling2011modeling},
\begin{linenomath*}
\begin{eqnarray}
\frac{dS}{dt}&=&-\left(\beta \frac{S}{N_o}\right) I, \\ \nonumber
\frac{dI}{dt}&=&\left(\beta \frac{S}{N_o}\right) I - \gamma I, \\ \nonumber
\frac{dR}{dt}&=&\gamma I,
\label{eq:SIRM}
\end{eqnarray}
\end{linenomath*}
where $t$ is time, $\beta$ is an average number of contacts per individual per unit time, and $\gamma^{-1}$ is the average time period required for an infected individual to get recovered. The $\beta S/N_o$ describes the force of infection \citep{dietz2002daniel, cadoni2020size} and coefficients $\beta$ and $\gamma$ are externally supplied constants that depend on the pathogen, environmental conditions, and the behaviour of the infected population (discussed later).  This system requires as initial conditions $I(0)$ to be finite, $S(0)=N_o-I(0)$ and $R(0)=0$ at the initial phase of the outbreak.  By summing the three equations describing the SIR system yields $d(S+I+R)/dt=0$, and then upon integration, result in $S(t)+I(t)+R(t)=N_o$ being constant at all times.  The constant $N_o$ can be evaluated from the initial condition ($t=0$). 

The SIR model makes several assumptions that include no large imbalances in natural births or natural deaths during the short-lived outbreak.  Direct transmission occurs through individual-to-individual contact represented by a coefficient of $\beta$.  The infection is assumed to have a zero latent period so that an individual becomes infectious as soon as they become infected.  However, the most restrictive assumption in SIR dynamics is the mass action of individuals. Mass action law assumes that the rate of an encounter between $I$ and $S$ is proportional to their product. This requires that infected and susceptible individuals be uniformly distributed within the space of the box.  It may be argued that for high $I$ and low $S$ densities, the mass action law must be revised to accommodate saturation effects due to competition for space (as infected individuals become reasonably dense, most infected individuals are surrounded by infected individuals thus having lower possibility of interacting with the susceptible individuals).  This saturation or second-order effect is ignored here.  The $\gamma$ and $\beta$ encode the main properties of the epidemics and the population response to it and will be assumed constant in time and space.

For illustration purposes only, we use values for $\gamma$ based on those used in models of the dynamics of the recent COVID-19 pandemic i.e. the time from the onset of symptoms to natural recovery is, on average, set to $\gamma^{-1}=14$ d \citep{katul2020covid}.  To be clear, the goal is not to explore any particular spreading event of COVID-19 for a problem-specific setting. The choice of a $\gamma^{-1}$ being of order 10 days is selected because geographical mobility usually occurs on time scales much faster than $\gamma^{-1}$.  The remaining model parameter $\beta$ must be determined empirically or from separate studies, which can pose difficulties \citep{keeling2009mathematical, turkyilmazoglu2021explicit}.   It is evident that $dI/dt$ will be positive (outbreak) or negative (epidemic contained) depending on the sign of $(\beta S/N_o-\gamma)$.  Defining the basic reproduction number $R_o=\beta/\gamma$ and assuming $S\approx N_o$ at early times (i.e. $I(0)/N_o\ll1$) results in the necessary condition for an outbreak: $R_o>1$.   We track the sensitivity of front speeds and spatial pattern formation of infections on $R_o$ while setting $\gamma^{-1}$ constant to 14 days.  Typical values of $R_o$ have been reported for a number of diseases and range between 2 to 8 (including COVID-19 \citep{katul2020covid}).

Spatially, $\gamma$ and $R_o$ are assumed constant for each box during an outbreak though effectiveness of certain treatment can reduce the average time between onset of symptoms and recovery. Likewise, a constant $R_o$ and $\gamma$ ignores other factors such as stay-at home orders, masking, social distancing, hand washing, etc....In general, these two parameters may also vary with other environmental or hydro-climatic or seasonal conditions \citep{keeling2001seasonally, hooshyar2020cyclic} not covered and are outside the scope of the work here. 

\subsection{A Spatial Model with Non-local Mobility of Infectious Individuals}
In a closed system which we label as box, the SIR model prohibits mobility of individuals across the box boundaries.  Mathematical approaches extending the SIR model from a single box (that may be experiencing mass action mixing) into a Cartesian ($x$,$y$) plane represent the spatial domain by a square-lattice with characteristic size $L$ and area $A_d=L^2$.  This domain is then filled with individual boxes of dimensions $dx$ by $dy$ with each box representing a community characterized by an initial population $N_o$.  The center of an arbitrary box is defined by its coordinates $x,y$. The basic SIR dynamics (i.e. conversion of $S$ to $I$ to $R$) along with mass action only applies within the box positioned at $(x,y)$. The size of each box is assumed to be much smaller than the domain size to allow continuous treatment of space.  Each box is subject to the initial condition $S(0)+I(0)+R(0)=N_o$. 

\begin{figure}
\centerline{\includegraphics [angle=0,height=0.51\linewidth]{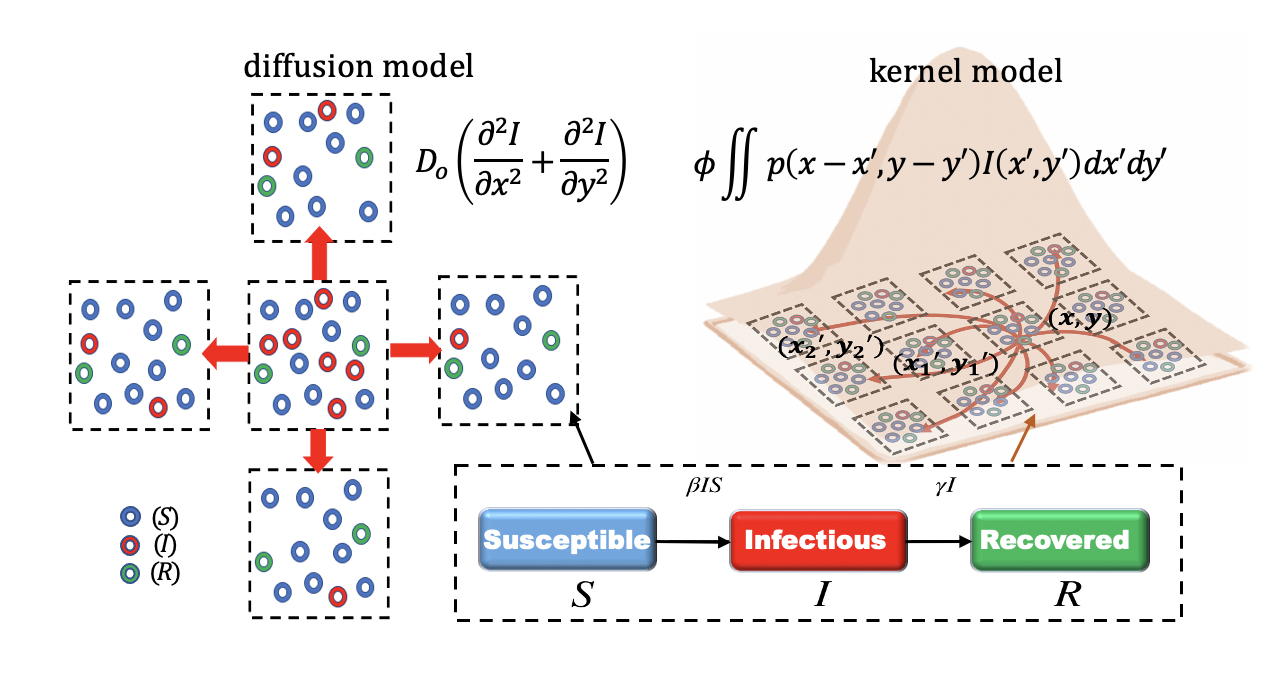}}
\caption{A conceptual diagram of the SIR model and its extension in space. In each box (dashed square), there are three compartments ($S$, $I$, and $R$) describing the composition of its population. Each box can be understood as an a community abiding by SIR dynamics when the infectious individual(s) arrive in this box. To describe the interaction between any two communities, two different transport models are presented. The left shows spatial spread across communities (or boxes) represented by a diffusion scheme with a diffusion coefficient $D_o$, while the right plot shows the spread between communities using the kernel-based model accounting for the non-local effects (i.e. infectious individuals can access all points in the spatial domain).  In the diffusion case, only spatial gradients in $I$ drive spatial mobility whereas in the kernel-based case, few infectious individuals can travel far from their parent community and access susceptibles in other communities or boxes.  These individuals are labelled here as geographical spreaders and constitute a small fraction ($\phi$) of the infected population.}
\label{fig:sir} 
\end{figure}

The concern here is the spatial mobility of $I$ across boxes (geographical spreaders) and its subsequent spatial patterns when few infectious individuals randomly occur in few boxes within the domain.  Because infectious individuals can be mobile, conventional spatial models revise the original SIR to allow for their mobility using a diffusion term given as  
\begin{linenomath*}
\begin{equation}
\frac{\partial I}{\partial t}=\left(\beta \frac{S}{N_o}-\gamma\right) I -D_o \left(\frac{\partial^2 I}{\partial x^2}+ \frac{\partial ^2 I}{\partial y^2}\right),
\label{eq:SIRDo}
\end{equation}
\end{linenomath*}
where $D_o$ is a diffusion coefficient that must be a priori known. This representation is continuous in space and time, meaning $dx$ and $dy$ are infinitely small - at least when compared to the domain size. Such systems are known to admit travelling fronts whose stationary speed $c_{max}$ is, to a leading order, given by the early times dynamics (when $S\approx N_o$) \citep{fisher1937wave,volpert2009reaction,murray2001mathematical, belik2011natural} 
\begin{linenomath*}
\begin{equation}
\label{eq:c_max_diff}
c_{max} =2 \sqrt{D_o \gamma (R_o-1)},
\end{equation}
\end{linenomath*}
when $R_o>1$.  This relation between $c_{max}$ and $\sqrt{D_o \gamma (R_o-1)}$ can also be derived from dimensional considerations alone. It is to be noted that equation \ref{eq:c_max_diff} requires $(R_o-1)>0$, which agrees with the standard SIR model without a spatial component where $R_o>1$ was necessary for an epidemic to form. To allow for non-local mobility of geographical spreaders, we formulate the \textit{I} budget as
\begin{linenomath*}
\begin{equation}
\frac{\partial I}{\partial t}= \left(\beta \frac{S}{N_o}-\gamma\right) \left[(1 -\phi) I + \phi C_u(x,y)\right],
\label{eq:SIRK}
\end{equation}
\end{linenomath*}
with the convolution function $C_u(x,y)$ defined as
\begin{linenomath*}
\begin{equation}
C_u(x,y)=\int_{-\infty}^{+\infty}\int_{-\infty}^{+\infty} I(x',y') p(x-x', y-y') dx' dy'
\label{eq:SIRmb}
\end{equation}
\end{linenomath*}
where $\phi$ is the fraction of infectious individuals (geographical spreaders) that are mobile across boxes within a given time step $dt\ll\gamma^{-1}$ and $p(x',y')$ is a spatial spread kernel satisfying the normalizing condition
\begin{linenomath*}
\begin{equation}
\int_{-\infty}^{+\infty}\int_{-\infty}^{+\infty} p(x', y') dx' dy'=1.
\label{eq:SIRa}
\end{equation}
\end{linenomath*}
The conceptual difference between the two models in equations \ref{eq:SIRDo} and \ref{eq:SIRK} is explained in Figure \ref{fig:sir}.

\noindent When $\phi=0$, the IDE approach reduces to spatially independent or autonomous SIR models operating in closed communities with no connectivity or spatial interaction between boxes (i.e. immobile model). When $\phi=1$, all infectious individuals experience spatial mobility per unit time. The fact that the spread kernel acts directly on $I$ is in keeping with recent findings derived from network theory \citep{brockmann2013hidden} that suggested the number of infectious individuals is more significant than effective distances between connected areas.  

Several objections can be raised when setting $\phi$ to a finite constant. The first is that $N_o$ is no longer a constant within a given box. The second is that spatial mobility is exclusive to $I$ not $S$ and $R$, which is not realistic.  However, a recent analysis on a similar IDE model has demonstrated that the spatial mobility of $I$ is far more critical to disease spread than the spatial mobility of $S$ and $R$ \citep{Genge2023321118}.  To arrive at key analytical results, the analysis is restricted to small $\phi(\ll1)$ values to illustrate how few geographical spreaders impact pattern formation and infectious fronts.  

There are a number of choices that can be made about $p(x',y')$ \citep{jo2014analytically, gleeson2016effects, keeling2011modeling}. For simplicity and to recover $c_{max}$ predicted by diffusion, we selected  
\begin{linenomath*}
\begin{equation}
p(r)=\frac{\alpha_N}{\sqrt{2\pi}\sigma} \exp\left[-\frac{1}{2}\left(\frac{r}{\sigma}\right)^2\right],
\label{eq:SIRpr}
\end{equation}
\end{linenomath*}
where $r=\sqrt{(x-x_o)^2+(y-y_o)^2}$ is the distance, $(x,y)$ and $(x_o,y_o)$ are two points in the domain representing the box centers, $\sigma$ is a measure of the spread of the spatial kernel (to be related to $D_o$), and $\alpha_N$ is a normalizing constant.  Since the interest here is in spatial spread kernels with finite support $R_a$ set by $dr<R_a<L$ where $dr^2=(dx)^2+(dy)^2$, $\alpha_N$ is determined so that \citep{fuentes2003nonlocal,fuentes2004analytical,da2009self}
\begin{linenomath*}
\begin{equation}
\frac{\alpha_N}{\sqrt{2\pi}\sigma} \int_{-R_a/\sqrt{2}}^{R_a/\sqrt{2}}\exp\left[-\frac{1}{2}\left(\frac{r'}{\sigma}\right)^2\right]dr'=1.
\label{eq:SIR6}
\end{equation}
\end{linenomath*}
This condition yields the normalizing constant
\begin{linenomath*}
\begin{equation}
\alpha_N(R_a,\sigma)=\left[\rm{erf} \left(\frac{R_a}{2 \sigma} \right)\right]^{-1}.
\label{eq:SIR7}
\end{equation}
\end{linenomath*}
To maintain a near-Gaussian shape for the spreading kernel, we selected $R_a/(2 \sigma) > 1$. This IDE approach along with a Gaussian spreading kernel was recently employed to explain the multi-fractal properties of COVID-19 infections from scales spanning 10 km to 2800 km across the U.S. \citep{Genge2023321118} and from 1 week to a 1 year period when using $\gamma^{-1}=14$ d.  Other spatial kernels can also be specified and subjected to the same normalizing conditions thereby making the IDE approach flexible in terms of choices about spatial spread of infectious individuals across communities.  

\subsection{Basic Properties of the IDE Model}
The IDE system exhibits two properties relevant to the objectives here and are derived in 1-D chosen along $x$ for illustration.  The first is the existence of travelling waves and the determination of their speed.  The second considers necessary (but not sufficient) conditions for pattern formation using stability analysis around the only equilibrium point ($I=0$).  Both properties are analyzed in a 1-D case to show-case expected connections between these properties, $\phi$, the spatial kernel shape, and epidemiological constants $R_o$ and $\gamma$. The derivation is also linearized around the early times (i.e. infections are increasing exponentially) when $S(x,t)\approx S(x,0)$ so that the IDE can be reduced to
\begin{linenomath*}
\begin{equation}
\frac{\partial I}{\partial t} \approx \frac{\gamma} {\left(R_o-1\right )}\left[(1 -\phi) I + \phi C_u(x)\right].
\label{eq:SIR1d}
\end{equation}
\end{linenomath*}

The convolution between $I$ and $p$ in Equation \ref{eq:SIR1d} is now simplified as
\begin{linenomath*}
\begin{equation}
C_u(x)=\int_{-\infty}^{+\infty} I(x') p(x-x')dx'= I(x') \star p(x').
\label{eq:CovIp}
\end{equation}
\end{linenomath*}

\subsubsection{Travelling waves and front speed}
For establishing links between the diffusion representation leading to equation \ref{eq:c_max_diff} and the 1-D IDE system, the travelling wave speed is now considered. A variable transformation $z=x-ct$ along with $dx'=dz'$ is used to convert the 1-D IDE in equation \ref{eq:SIR1d} to the new coordinate system, given as
\begin{linenomath*}
\begin{equation}
-c\frac{\partial I}{\partial z} \approx \gamma \left(R_o-1\right )\left[(1-\phi)I +\phi C_u(z) \right].
\label{eq:SIR2f}
\end{equation}
\end{linenomath*}

\noindent Assuming the solution is of the form $I(z)=A\exp(\lambda z)$ and using a zero-mean Gaussian kernel $p(z)$, two Fourier functions can be used to aid in the evaluation of $C_u(z)$ and are defined as
\begin{linenomath*}
\begin{equation}
\hat{I}(k)= 2A\pi \delta(\lambda i +k),~ 
\hat{p}(k)= \exp\left(-\frac{1}{2}\sigma^2 k^2\right),
\label{eq:SIRp}
\end{equation}
\end{linenomath*}
where $A$ is a constant, $k$ is the wavenumber, $i^2=-1$,  $\delta(.)$ is the Dirac-delta function, and $\hat{s}$ denotes the corresponding Fourier transform of function $s(z)$.  To evaluate $C_u(z)$, the product of $\hat{I}(k)$ and $\hat{p}(k)$ is determined in the Fourier domain followed by an inverse Fourier transform yielding $C_u(z)=I(z) M_g(\lambda)$ where
\begin{linenomath*}
\begin{eqnarray}
M_g(\lambda)=\exp\left(\frac{1}{2}\sigma^2 \lambda^2\right)
\label{eq:SIRp1}
\end{eqnarray}
\end{linenomath*}
\noindent defines the moment-generating function of a zero-mean Gaussian kernel with variance $\sigma^2$.  Substituting this estimate of $C_u(z)$ into equation \ref{eq:SIR2f} and upon further simplification, a characteristic equation for $\lambda$ is derived as
\begin{linenomath*}
\begin{equation}
-c\lambda = \gamma \left(R_o-1\right )\left[(1-\phi) +\phi M_g(\lambda)\right].
\label{eq:SIRlam}
\end{equation}
\end{linenomath*}
The interest here is in monotonic (rather than oscillatory)  wave fronts and real (rather than complex) roots. Real roots
emerge as multiple roots at the nth-order contact  \citep{kot1996dispersal,thompson2008plant,volpert2009reaction,tian2017traveling} given by differentiating equation \ref{eq:SIRlam} with respect to $\lambda$.  Noting that $dM_g/d\lambda=\lambda \sigma^2 M_g$ and differentiating equation \ref{eq:SIRlam} with respect to $\lambda$ yields
\begin{linenomath*}
\begin{eqnarray}
-c = \gamma \left(R_o-1\right ) \lambda\left[\phi  \sigma^2 M_g(\lambda)  \right],
\label{eq:SIR5}
\end{eqnarray}
\end{linenomath*}
or as
\begin{linenomath*}
\begin{eqnarray}
\phi M_g(\lambda)=-\frac{c}{\lambda}\left[\frac{1}{\gamma \left(R_o-1\right )\sigma^2}\right]. 
\end{eqnarray}
\end{linenomath*}

When $R_o>1$ (outbreak), equation \ref{eq:SIR5} requires that $c/\lambda<0$ (i.e. $c$ and $\lambda$ must have opposite signs reflecting expansion and contraction phases of the disease spread).   Combining equations \ref{eq:SIRlam} and \ref{eq:SIR5} yields a $c$ given by 
\begin{linenomath*}
\begin{equation}
-c \lambda^2 -\gamma (R_o-1)(1-\phi) \lambda +\frac{c}{\sigma^2}=0, 
\label{eq:discrem}
\end{equation}
\end{linenomath*}
whose solution is 
\begin{linenomath*}
\begin{equation}
c_{max}=\frac{c_k} {1-c_k^2}\gamma\sigma(1-\phi)(R_o - 1), 
\label{eq:SIR14}
\end{equation}
\end{linenomath*}
where $c_k=\lambda\sigma$ is a control factor that varies with the initial conditions.  This solution admits two traveling wave velocities (i.e. two $\lambda$ values) depending on the initial and boundary conditions imposed on the IDE:  a forward velocity (when $\lambda<0$) and a backward velocity (when $\lambda>0$).  Conventional approaches to by-pass $c_k$ and determine a single $c_{max}$ assume that the discriminant of equation \ref{eq:discrem} with respect to $\lambda$ is zero to ensure double roots \citep{kot1996dispersal}.  In this application, the discriminant $\Delta$ is given by
\begin{linenomath*}
\begin{equation}
\Delta=\left[\gamma(1-\phi)(R_o - 1)\right]^2+\left[\frac{2 c_{max}}{\sigma}\right]^2>0, 
\label{eq:SIS_Discriminant}
\end{equation}
\end{linenomath*}
and demonstrates that no double root exists (i.e. disease expansion and contraction fronts must both occur).  

When $\phi\ll1$, as is the case here, $c_{max}$ appears to be insensitive to $\phi$ though $1-\phi$ does act as a retardation factor to the attainment of $c_{max}$ (discussed later). Unlike the diffusion model, the kernel-based $c_{max}$ varies linearly with $\gamma (R_o-1)$ instead of $\sqrt{\gamma (R_o-1)}$. The $c_{max}$ also scales linearly with the kernel spread measure $\sigma$. Equating the two speeds from equations \ref{eq:c_max_diff} and \ref{eq:SIR14} yields an effective $D_o$,
\begin{linenomath*}
\begin{equation}
\frac{D_o}{\gamma\sigma^2}=\frac{c_k^2}{4(1-c_k^2)^2}{(1-\phi)^2(R_o-1)}.
\label{eq:SIR15}
\end{equation}
\end{linenomath*}
Thus, diffusion models can be made analogous to the IDE approach only for the purposes of estimating stationary front speeds.  The variables in equation \ref{eq:SIR14} are derived here mainly to guide the numerical model analysis for invasion front speeds (in 2D) and for differing initial conditions for disease attack.   The analytical solution in 1-D here must only be viewed as informing a scaling relation between travelling wave speeds, epidemiological parameters, and kernel spread properties for more complex initial conditions and in a higher dimension. 

\subsubsection{Necessary Conditions for Pattern Formation}
As common to pattern formation studies, linear stability analysis is conducted by perturbing the disease free equilibrium state $I_e(x,t)$.  Near this equilibrium, a small perturbation is introduced so that $I(x,t)=I_e(x,t)+\epsilon\cos(\omega x) \exp(\varphi t)$, where $\epsilon$ is the amplitude of the small perturbation \citep{fuentes2004analytical,kefi2008vegetation,murray2001mathematical}.  Inserting $I(x,t)$ into equation \ref{eq:SIR1d} and noting that at equilibrium $\partial I_e(x,t)/\partial t=0$ yields
\begin{linenomath*}
\begin{equation}
\varphi=\left(R_o-1\right )\left[(1 -\phi)  + \phi \frac{h(x)}{\cos(\omega x)}\right],
\label{eq:Coherent1}
\end{equation}
\end{linenomath*}
where $h(x)=g \star p (x)$ is the convolution of $g=\cos(\omega x)$ and $p(x)$. To evaluate $h(x)$, recall that the Fourier transforms of $g$ (in $k$ or wave space) is
\begin{linenomath*}
\begin{eqnarray}
\hat{g}(k)&=&\sqrt{\frac{\pi}{2}}\left[ \delta(\omega-k) + \delta (\omega+k) \right]
\label{eq:Coherent2}
\end{eqnarray}
\end{linenomath*}
Upon computing the product of $\hat{g}(k)$ and $\hat{p}(k)$ in Fourier space and inverse Fourier transforming yields
\begin{linenomath*}
\begin{equation}
h(x)=\cos(\omega x)   \exp\left(-\frac{1}{2} \sigma^2 \omega^2 \right).
\label{eq:Coherent3}
\end{equation}
\end{linenomath*}
Inserting this outcome from equation \ref{eq:Coherent3} into equation \ref{eq:Coherent1} yields
\begin{linenomath*}
\begin{equation}
\varphi=\left(R_o-1\right )\left[(1 -\phi)  + \phi \exp\left(-\frac{1}{2} \sigma^2 \omega^2 \right)\right].
\label{eq:Coherentp4}
\end{equation}
\end{linenomath*}
A necessary condition for pattern formation is that $\varphi>0$ \citep{murray2001mathematical,fuentes2004analytical,segal2013pattern}.  Since $R_o>1$ (outbreak), $\varphi$ is always positive. It may be stated that the IDE is de-stabilized by small-amplitude perturbations at any wavenumber (in 1-D), which is a necessary (but not sufficient) condition for pattern formation.

\section{Simulation Results}
The simulation results are presented to address the two main questions: (i) to what degree $c_{max}$ derived from a linearized analysis in 1-D captures the key state variables describing propagation fronts in 2-D for an isolated outbreak in the domain center (set at $x_o, y_o$). (ii) What spatial patterns are expected when the domain becomes finite and is attacked at multiple communities leading to interfering wavefronts? and (iii) what is the role of the kernel properties ($R_a$, $\sigma$) and geographical spreader fraction $\phi$ on spatial patterns or any switching between them in time.   

\subsection{Numerical Simulations and their Presentation}
All calculations are conducted on a square domain with size $L\times L$ comprising of $N\times N$ square boxes (or lattices) with $dx=dy=1$ of arbitrary spatial units. For $N$, three different values ($N$=512, 1024 and 2048) are used and compared to ensure that the patterns are not sensitive to the domain size. The boundary conditions imposed on equation \ref{eq:SIR1d} are periodic. In such calculations, time $t$ is made dimensionless using $\gamma$.  Time-integration employs forward differences with dimensionless time increments $\gamma dt=0.1$. Spatial integration is conducted using two-dimensional Fast Fourier Transform (FFT) convolutions between $I(x',y',t)$ and $p(x-x',y-y')$. Variations in $R_a/L$ and $\sigma/R_a$ are explored with certain constraints set by (i) the smallest size $dr=\sqrt{dx^2+dy^2}=\sqrt{2}$ (arbitrary spatial units) and (ii) the largest size determined by the number of grid nodes $N^2=(L/dx) \times (L/dy)=2 (L/dr)^2$.  

Other disease-related parameters are also varied within certain ranges including $R_o$=2.2-8.0 reported for COVID-19 \citep{katul2020covid, kucharski2020early}, and the fractional mobility of the infected population ($\phi$=0.01 and 0.1). The latter is needed to assess whether coherent pattern shapes in infections are driven by finite bounds of the spatial domain.  Time evolution of spatially-averaged quantities are also required to highlight a number of features about the domain-averaged SIR system.  The domain-averaged quantity for an arbitrary variable $\Psi(x,y,t)$ is indicated by
\begin{linenomath*}
\begin{equation}
\langle\Psi\rangle(t)=\frac{1}{L^2}\int_{-L/2}^{L/2} \int_{-L/2}^{L/2}\Psi(x,y,t)dx dy.
\label{eq:Space_average}
\end{equation}
\end{linenomath*}
For notational simplicity, the time dependence is dropped and $\langle \Psi \rangle$ is assumed to represent the time- dependent quantity $\langle \Psi \rangle(t)$.

\subsection{Maximal Expansion and Contraction Speeds}
\begin{figure} %[!ht] 
\centerline{\includegraphics [angle=0,height=0.63\linewidth]{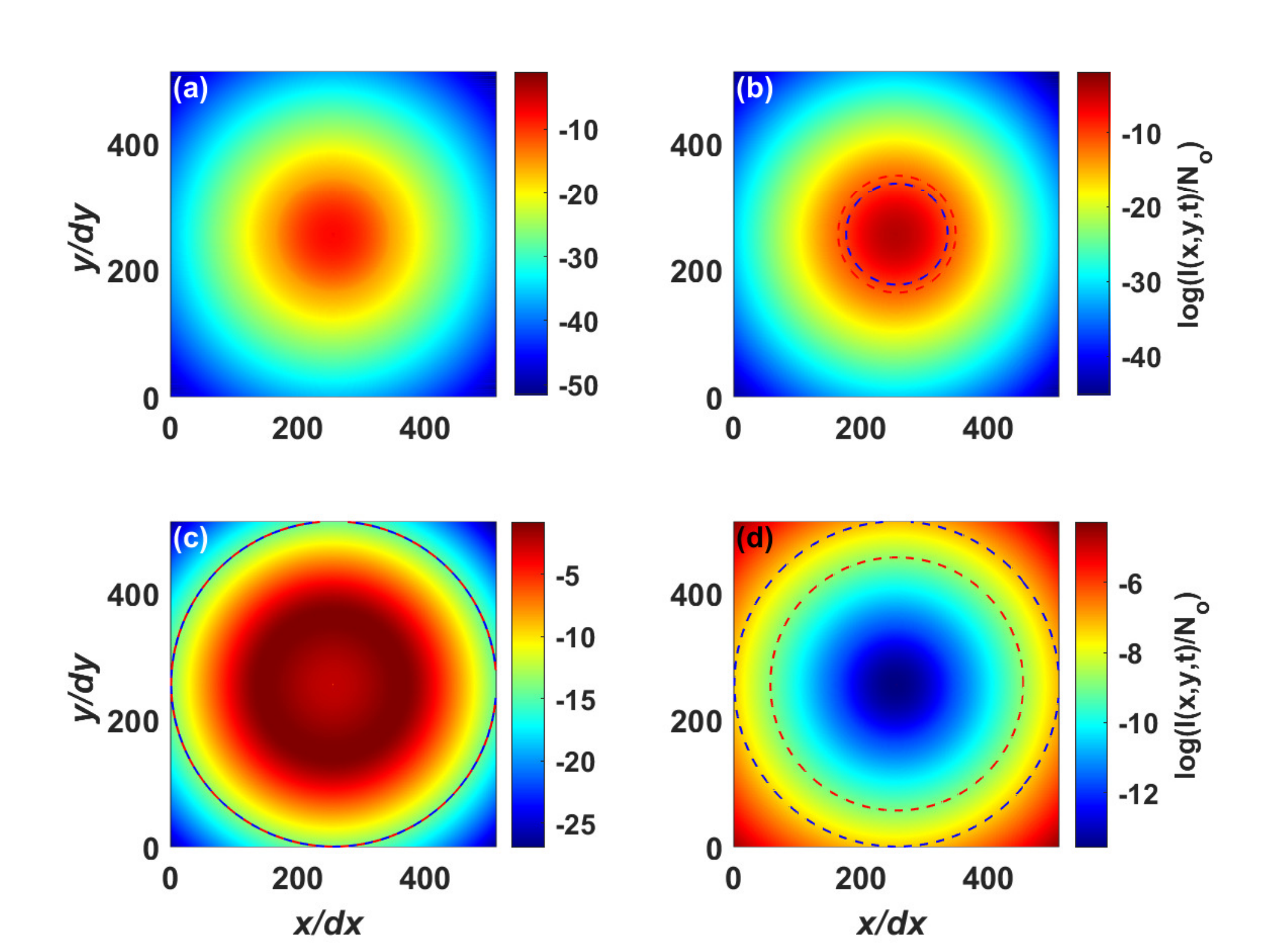}}
\caption{Representative contours of the spreading area $A_s(t)$ and the maximal expansion and contraction speeds for an initial infection at the center of the domain ($I(256,256,0)=50$). The blue and red circles outline the epidemiological expansion area directly captured by the maximal spatial gradient and by $A_s(t)$, respectively. In all the plots $\phi=0.01$, $R_o=4.5$, $R_a/dr=64$, $\sigma/dr=16$, and $\gamma t$=2, 3, 7 and 18.6 respectively.}
\label{fig:speed} 
\end{figure}

\begin{figure*} %[!ht] 
\centerline{\includegraphics [angle=0,height=0.6\linewidth]{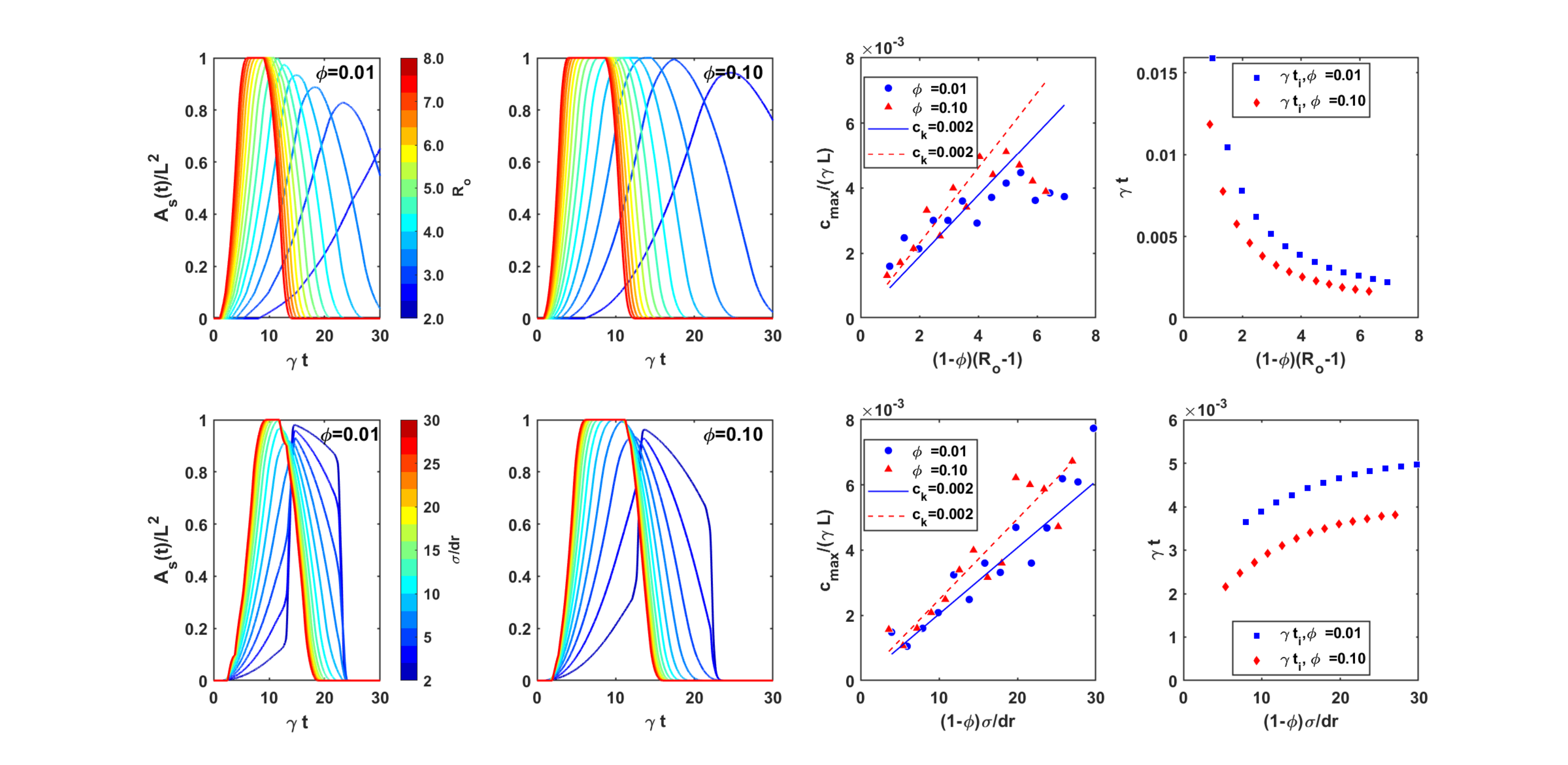}}
\caption{Variation of the fractional spreading area $A_s(t)$ versus dimensionless time and the maximal expansion speeds $c_{max}$ for an initial infection at the center of the domain. The blue and red dashed lines represent the analyzed velocity using Equation \ref{eq:SIR14}. In all plots, $R_a/dr=64$, $\sigma/dr=16$ in the top panel and $R_o$=4.5 in the bottom panel.}
\label{fig:speed1} 
\end{figure*}

Before presenting the pattern formation results for multiple attack points as initial conditions, the $c_{max}$ are first discussed for different kernel properties and $R_o$ with idealized initial conditions.
These conditions feature a concentrated initial infectious event of 50 individuals positioned at the center of the domain in one box. The spreading area $A_s(t)$ is used throughout to delineate overall spreading within the entire domain.  It is computed from the fractional area where $I(x,y,t)/N_o>0.1\%$. While this threshold appears arbitrary, it is selected as a compromise between the need for front detection and a sufficient number of infections to initiate early dynamics in SIR locally within a box.  This allows estimates of a spatially-integrated front speed $c(t)=d\sqrt{A_s(t)}/dt$ (expansion when $dA_s/dt>0$ and contraction when $dA_s/dt<0$) when multiple infectious points produce interfering front waves in later analyses.  A simulation result featuring the idealized initial condition and the computed area expansion and contraction as well as front speed are shown in Figures \ref{fig:speed} and \ref{fig:speed1}.  This front detection is compared with another approach based on maximal spatial gradients in $I$ to ascertain that the results derived from a spatially integrated measure such as $A_s$ is comparable to standard gradient based on estimates of fronts.

In Figure \ref{fig:speed}, the expansion area calculated with $A_s(t)$ captures the front movement indicated by the comparison of white and red circles lending confidence to the use of $A_s(t)$ for $c_{max}$ detection. In Figure \ref{fig:speed1}, the numerical model results show that $ c_{max}$ is indeed proportional to $(1-\phi)(R_o-1)$ when setting $\sigma$ constant and to $\sigma(1-\phi)$ when setting $R_o$ to a constant, which agrees with the 1-D analysis. More significant, the $c_{max}$ values do not vary much even when $\phi$ is increased from 0.01 to 0.10, implying that the fraction of geographical spreaders in the infectious population has a minor impact on the magnitude of the maximal spread velocity attained.  However, Figure \ref{fig:speed} also demonstrates that $\phi$ dictates the time at which $c_{max}$ is reached.  Thus, $\phi$ is pertinent for identifying 'early-warning' signals about impending outbreaks.  

Figure \ref{fig:speed1} suggests that for the values of $\phi$ and $\sigma$ explored here, a front cannot form and persist when $R_o<3$. This implies that $R_o-1>1$ in all model calculations to be conducted later. Interestingly, the 1-D model reasonably captures the 2D simulation results. However, a saturation in front speeds begins to emerge at very large $R_o>7$ irrespective of $\phi$, which cannot be predicted from the 1-D linearized analysis. This saturation in $c_{max}$ with increasing $R_o$ arises because of a rapid areal expansion resulting in a jump between no infections to an all-domain infection in a short time interval difficult to fully resolve with finite $dA_s/dt$. When the initial condition is fixed, the fitted constant $c_k$ to modeled $c_{max}$ does not change suggesting it is not sensitive to the kernel properties or $\phi$.  When $\sigma/dr<3$, kernel shapes are too narrow where much of the $I$ spreading occurs within an individual box instead of across boxes. This mode of spreading violates the mass action assumption representing interactions between $I$, $S$, and $R$ within the box. Therefore, a constraint on the kernel standard deviation $\sigma/dr>2$ is imposed in all later calculations.  This constraint is above and beyond the prior constraint $R_a/\sigma>2$ to ensure a near-Gaussian shape.  
When $3<\sigma/dr<30$, a linear relation between $c_{max}$ and $\sigma$ is confirmed by the simulations as conjectured from equation \ref{eq:SIR14}. Hence, we imposed $R_o>2$, $\sigma/dr>2$, and $R_a/\sigma>2$ in later numerical simulations where the domain is initially infected at multiple locations and pattern formation is tracked.

%An anomalous behavior at $\sigma/dr<2$ is observed in the simulation results here. This anomalous behavior may have been anticipated from equation \ref{eq:SIR14} when $\sigma$ is small as $c_{max}$ weakly scales with $\sigma^2$. This scaling arises due to $c_k=\lambda \sigma$ and thus ${c_k}\sigma/{(1-c_k^2)} \sim \sigma^2$ for small $\lambda \sigma$.

\subsection{Formation of Infectious Spatial Patterns}
\begin{figure*} %[ht] % 
\centerline{\includegraphics [angle=0,height=0.54\linewidth]{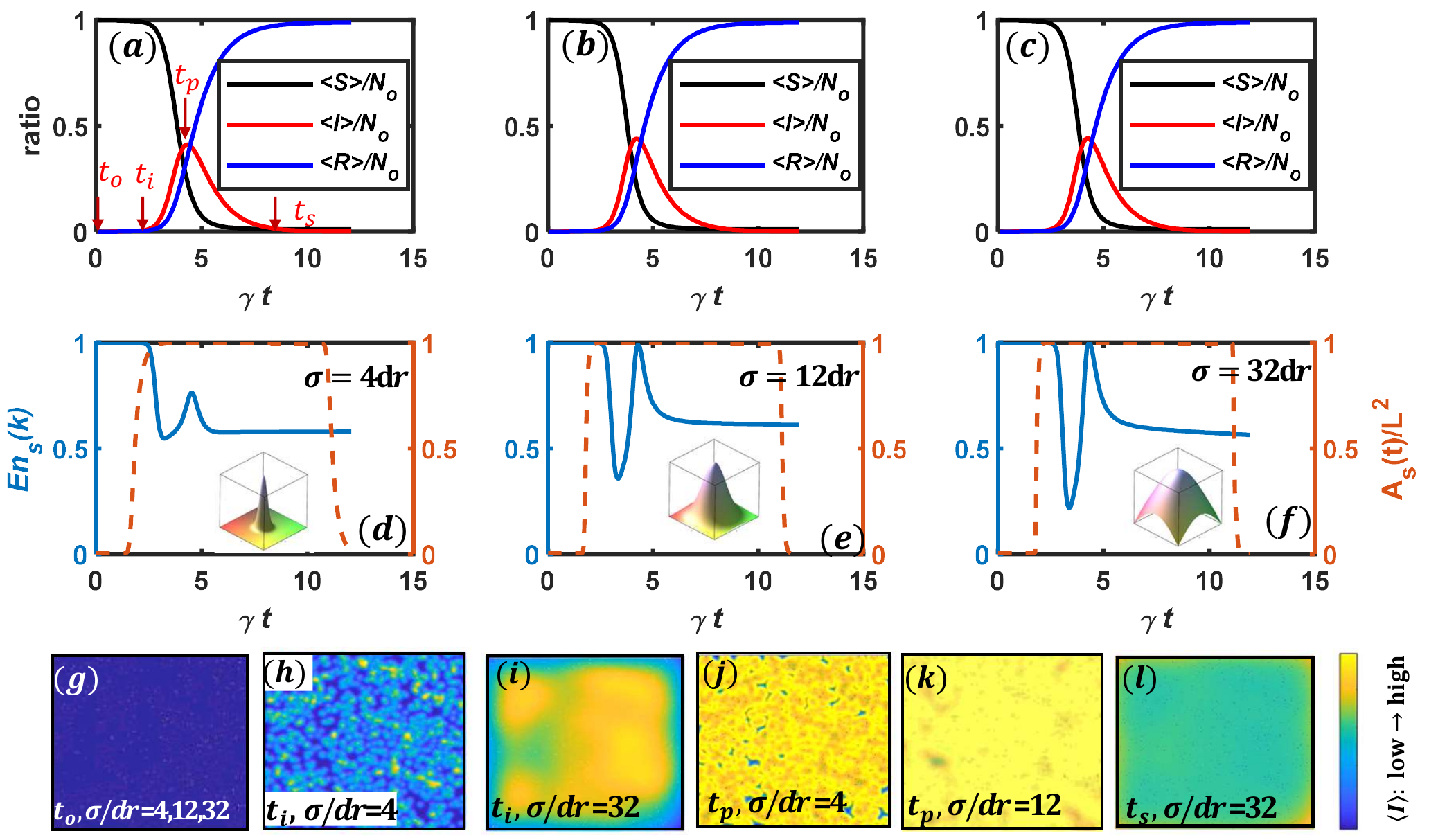}}
\caption{The variations of $\langle S \rangle$, $\langle I \rangle$, and $\langle R \rangle$ normalized by the total initial population ($N_o$ as a function of dimensionless time $\gamma t$). The solution starts at $t_0=0$ with spatially random infectious locations. The $t_i$ is the time when $d\langle I\rangle /dt$ reach its maximum, $t_p$ is the point when $\langle I\rangle$ becomes its maximum and $t_s$ is the quasi-steady state where infections are slowly approaching their zero equilibrium. The representative instances under consideration are labeled with red arrows along with the typical patterns expected. Plots (d) to (f) show the variation of entropy $En_s$ and spreading area $A_s$ over time, where the shapes of the kernels with different $\sigma$ are also sketched to illustrate the kernel spread within the domain. Plots (g) to (l) show the variation of the spatial structure of $I(x,y)$ depending on different kernel information ($\sigma/dr$) at the four characteristic times. In all the plots $R_a/dr=128$, $R_o=4.5$.}
\label{fig:sirt} 
\end{figure*}

\begin{figure*} %[ht] 
\centerline{\includegraphics [angle=0,height=0.54\linewidth]{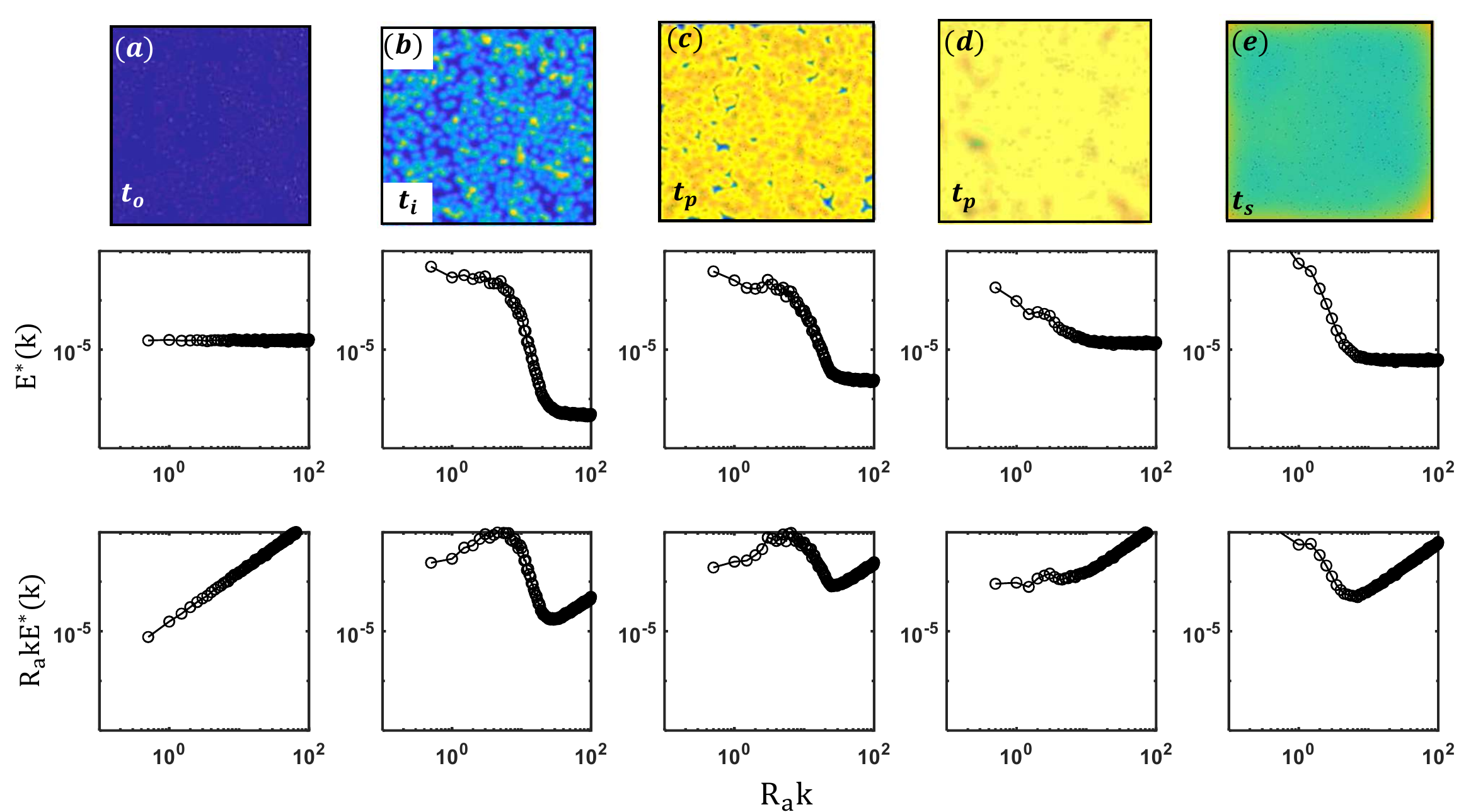}}
\caption{The first row shows spatial variations of $I(x,y,t)$ at key time instances $t_o$, $t_i$, $t_p$, and $t_s$ with different kernel parameters, which correspond to the patterns shown in Figure \ref{fig:sirt}. The second row shows the corresponding spectral density $E^*(k)$, while the third row shows the energy content $R_a k E^*(k)$. The wavenumbers are normalized by $R_a$, the kernel bound.  As evidenced by the third row, the so-called pre-multiplied spectral representation $k E^*(k)$ better delineates key scale transitions associated with changes in the spectrum.}
\label{fig:pt} 
\end{figure*}

In all the following simulations, 5\% of the communities delineated by boxes in the domain are randomly selected and simultaneously infected with 10 individuals as initial conditions.  Based on the IDE solution for $\langle I(x,y,t)\rangle $ versus $\gamma t$, the SIR model exhibits transient spatial patterns depending on choices made on $R_a$ and $\sigma$. Typical transient patterns of three structures are featured and analyzed in Figures \ref{fig:sirt} and \ref{fig:pt}. To interrogate a broad range of parameter space reflecting disease transmission ($R_o$) and kernel shape ($\sigma$ and $R_a$) on pattern formation, a dimensionless spectral Shannon entropy is proposed, given by
\begin{linenomath*}
\begin{equation}
En_s(t)=\frac{\sum_{i=1}^{i=m} E^*(k) \log[E^*(k)]}{\log (m)},
\label{eq:Shannon_Ent}
\end{equation}
\end{linenomath*}
where $m(=N/2)$ is the number of Fourier modes used in the computation of the Fourier energy spectrum, $E^*(k)$ is the squared amplitude at the scalar wavenumber $k$, and the normalization is the Shannon entropy for a white-noise spectrum determined at $m$ wavenumbers.  When $En_s=1$, the spatial energy spectrum of $I(x,y,t)$ is white and no patterning is expected (at t=0) whereas $En_s=0$ implies that the entire spatial variability is explained by a single Fourier mode (periodic function).  Thus, $En_s(t)$ is used here as a scalar measure whose time evolution determines how energy becomes concentrated among Fourier modes starting from an initial white-noise spectrum.  It is to be noted that the 1-D linearized analysis predicts that the IDE is unstable at all modes. However, such instability does not guarantee pattern formations.  The calculations of $En_s(t)$ are complemented by the constraint that energetic modes bounded in the interval $[5 dr, L/2]$ may be used in identifying coherent spatial patterns if this pattern arises (i.e. low $En_s$). This constraint is to ensure that any deterministic spatial patterns do not have wavelengths too close to the domain size or grid resolution. The temporal variation of the Shannon entropy is also featured in Figure \ref{fig:sirt} for illustration.

Figure \ref{fig:sirt} shows that $En_s(t)$ has distinct features at four times: (i) initial state ($t_o$), (ii) initial exponential rise in infections ($t_i$), (iii) peak infections ($t_p$) and (iv) the final quasi-steady state stage where infections are slowly approaching their zero-equilibrium value ($t_s$). When the spatial pattern of $I(x,y,0)$ is random in space, $En_s(t)$ is close to unity as expected at $t_o$. The structure at $t_i$ is transient due to the persistent impact from the random initial condition, while the emerging organized structure at $t_p$ can show different spatial patterns characterized by different $En_s(t)$ values. These differences frame much of the discussion about pattern formation. 

\begin{figure} %[ht] % 
\centerline{\includegraphics [angle=0,height=0.36\linewidth]{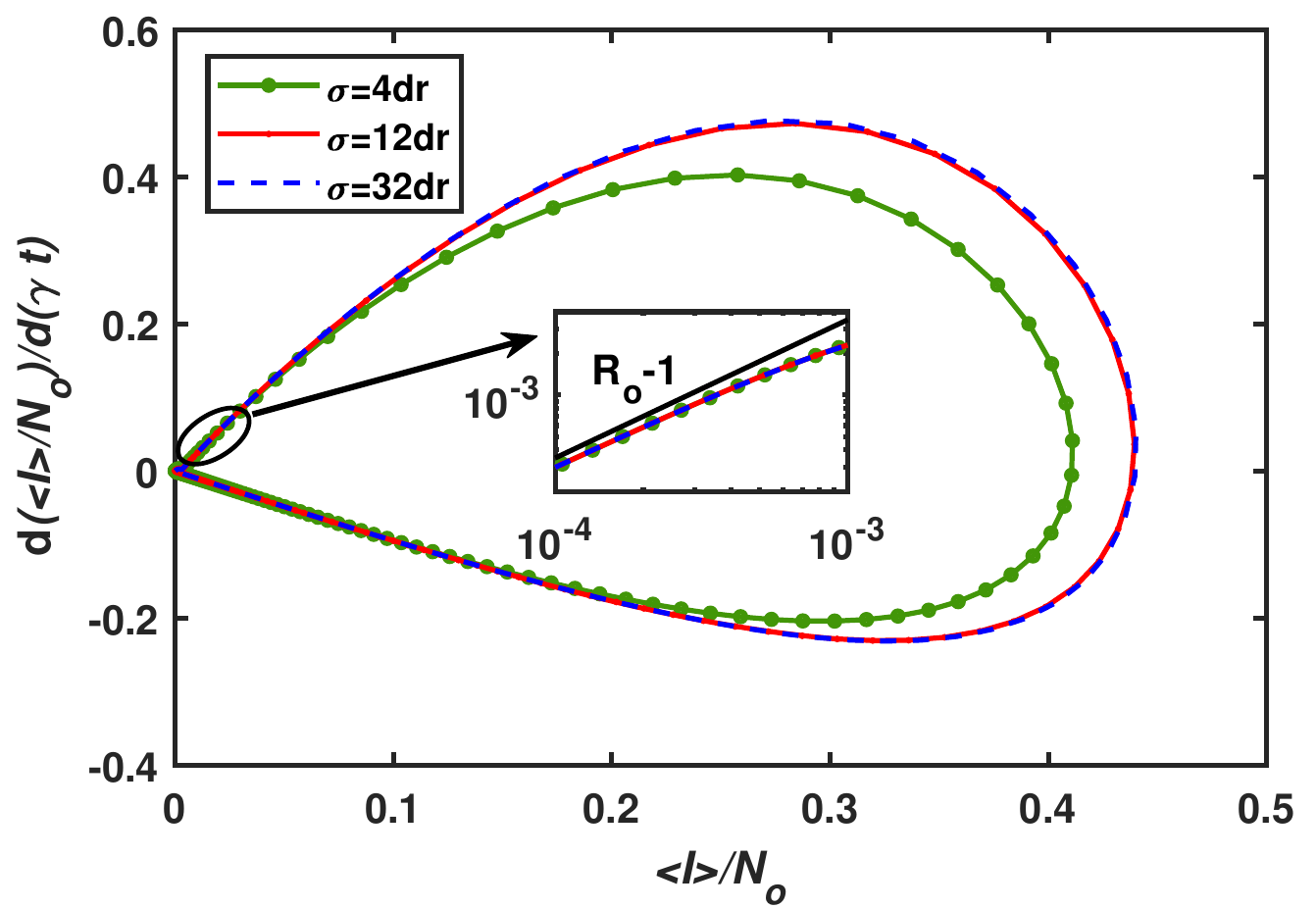}}
\caption{Phase space of the $N_o^{-1}d\left \langle I \right \rangle d(\gamma t)$ and $\left \langle I \right \rangle N_o^{-1}$. The inset is a zoom-in over the initial period where the black solid line is the expected ($R_o-1$) slope. }
\label{fig:sirph1} 
\end{figure}

Before delving into those differences, several points can be made about the dynamics of $\left \langle I \right \rangle$ first. First, it is clear that increases in $\sigma/dr$ has minor (i.e. graphically not discern-able) impact on the temporal patterns of $\left \langle I  \right \rangle$ (Figure \ref{fig:sirt}). However, exploration of the phase space of $d\left \langle I \right \rangle/dt$ versus $\left \langle I \right \rangle$ reveals a more nuanced perspective in Figure \ref{fig:sirph1}.

When $\sigma/dr$ increases, there is a clear shift in the phase-space until saturation at a large $\sigma/dr$ is attained. This finding hints that there may be a critical $\sigma$ for which disease spread becomes quasi-independent of $\sigma$ when measured by a spatially aggregated quantity such as $d\left\langle I \right \rangle/dt$.  For early-times SIR, all three $\sigma/dr$ collapse on an approximate line with a $(R_o-1)$ slope as discussed elsewhere \citep{katul2020covid}. However, as $t_p$ is gradually approached, $d\left \langle I \right \rangle/dt$ begins to decline from its maximum value, and the effects of $\sigma/dr$ become more evident.  During the expansion phase, higher $\sigma$ leads to higher $d\left \langle I \right \rangle/dt$ at a given $\left \langle I \right \rangle$.  Moreover, at $t_p$ where $d\left \langle I \right \rangle/dt=0$, the maximum $\left \langle I \right \rangle$ that can be supported mildly increases with increasing $\sigma$ but saturates to a $\left \langle I \right \rangle/N_o=0.45$.  It is to be noted that increasing $\sigma/dr$ by a factor of $8$ only leads to an increase in the maximum $\left \langle I \right \rangle/N_o$ at $t_p$ by some 12$\%$.   

Moving beyond the dynamics of $\left \langle I \right \rangle$, and depending on the kernel properties ($Ra$ and $\sigma$), the spatial patterns of $I(x,y,t_p)$ vary from disordered to organized (defined below). This spatial pattern shift at $t_p$ does not have an appreciable impact on $\left \langle I \right \rangle$ as earlier noted.  The analysis is now focused on the pattern formation of $I(x,y,t_p)$, namely, the variation of pattern topology exemplified in panels (h), (g), and (i). Corresponding to Figure \ref{fig:sirt}, the following descriptors are used on spatial patterns for clarity. These descriptors are aimed at qualitatively identifying key spatial patterning features in physical and spectral spaces (Figure \ref{fig:pt}). 

\textbf{i) Disordered Structure}: In physical space, these patterns are characterized by sharp interfaces that delineate infected from non-infected regions. In spectral space, this pattern can be discerned when an energetic mode with maximal spectral energy content (instead of spectral density) arises. This mode is detected at an intermediate wavenumber ($k_e$) where the spectral energy content $\sigma k E^*(k)$ reaches its maximum and obeys the constraint $k_o\ll k_e\ll k_r$ where $k_o \sim 1/L$ and $k_r \sim 1/dr$ are the limiting wavenumbers in Fourier space dictated by domain size $L$ and spatial resolution $dr$. Such a disordered pattern is shown in panels (b) and (c) of Figure \ref{fig:pt}.

\textbf{ii) Coherent Structure}: In physical space, these patterns emerge after the infectious population spreads across the entire spatial domain. The infections are space-filling and no lattice remains free from infections within the lattice. Thus, the $I(x,y,t)/N_o$ in space is primarily dominated by high values interspersed with infrequent but relatively smaller infectious spots of low (but finite) $I(x,y,t)/N_o$. In spectral space, a cluster of several contiguous dominant energetic modes with characteristic scales smaller than $L$ emerge.  However, the energy content $\sigma k E^*(k)$ of the large wavenumbers remains high and quasi-random when compared to wavenumbers commensurate with $k_o$. This feature is shown in (d) and (e) of Figure \ref{fig:pt}.

\textbf{iii) Pseudo-diffusive Structure}: In physical space, the transitions between infected and non-infected zones are almost obscured corresponding to the panel (i) in Figure \ref{fig:sirt}. The $I(x,y,t)/N_o$ in physical space spans numerous values. In spectral space, the pattern is associated with $k_e$ approaching $k_o$, meaning the spectral energy reaches its maximum at the lowest wavenumber (space-filling), and intermediate to high wavenumber modes are not energetic. The domain size clearly impacts $k_o$ here (as expected) - but the qualitative aspect of the pattern remains less affected by the domain size (i.e. as long as the pattern is almost space-filling irrespective of the size). Since it is a transient structure and can be impacted by domain size, it is not the main focus here. For completeness, its spectral properties are discussed in the Appendix for three domain sizes to illustrate the qualitative aspects of such pattern type and its sensitivity to domain size.

In all calculations, the dimensionless energy $E^*(k)$ at dimensionless time $\gamma t$ here is determined using a 2-D FFT of $I(x,y,t)$, computing the squared amplitudes (in $k_x$ and $k_y$ directions), normalizing by the overall variance, and setting the energy content as representative of wavenumber $k=\sqrt{k_x^2+k_y^2}$, $k=1/r_o$ is the wavelength and $dr<r_o<L$ is the spatial scale. 

Figure \ref{fig:pt} shows the initial condition (at $t_o$) that then set a white noise spectral pattern identical for all simulations. Likewise, the steady-state patterns (at $t_s$) are similar for all simulations.  At $t_i$, when $En_s$ begins its fastest decline in time (i.e. energy is being concentrated across scales), two possible spatial patterns emerge: diffusive (sensitive to domain-size) and disordered. In both cases, energy content begins to concentrate at low wavenumbers.  At $t_p$, the pattern forming structures are either disordered or coherent (both not sensitive to domain size). For both pattern types, disordered and coherent patterns have dominant energetic amplitudes at an intermediate scale that is smaller than the domain size but much larger than the grid resolution. The kernel properties that decide on which pattern persists at $t_p$ are to be investigated and discussed.   

\subsection{Impact of Kernel Size on Pattern Formation}
\begin{figure} %[ht] 
\centerline{\includegraphics [angle=0,height=0.48\linewidth]{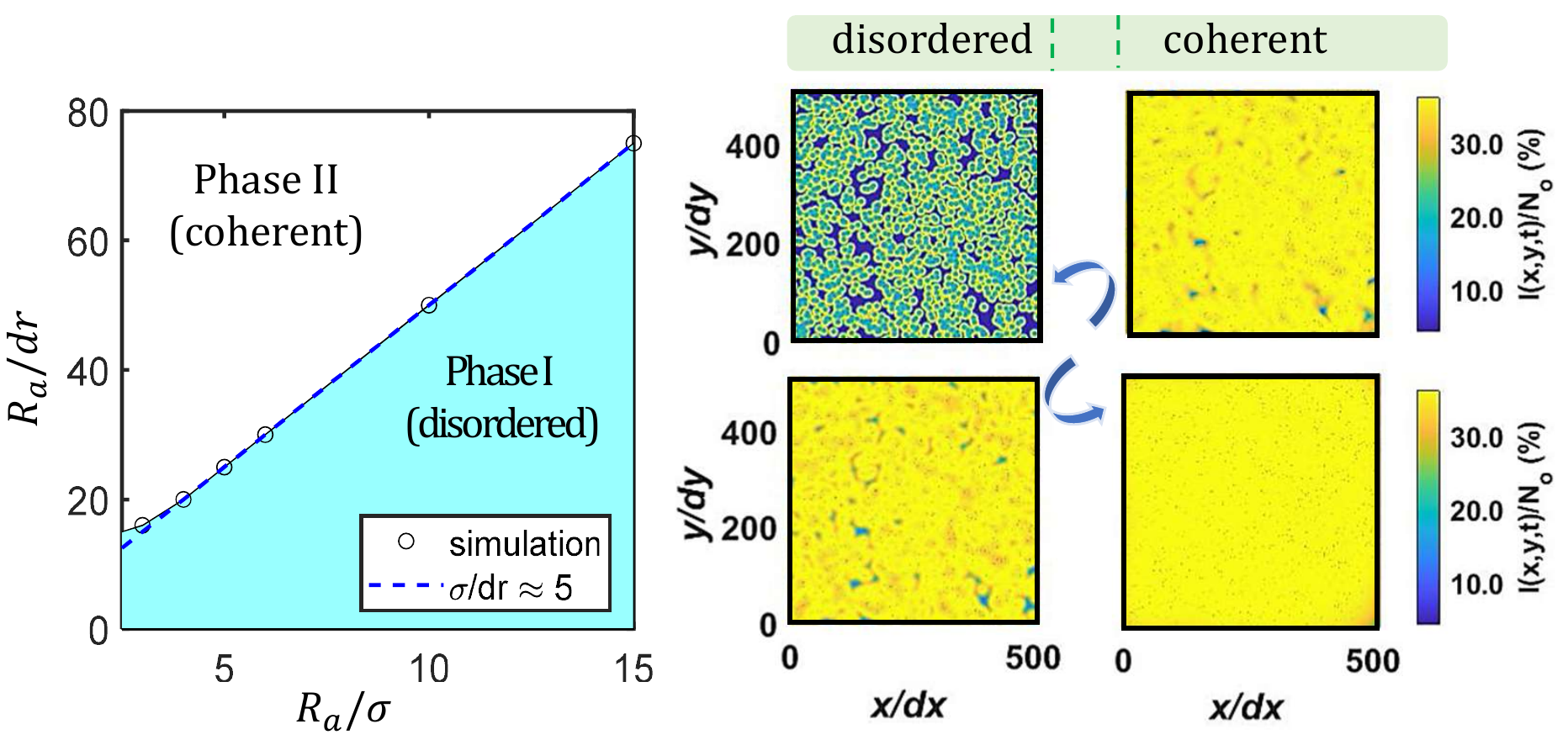}}
\caption{Spatial map for the pattern transition at $t_p$. The occurrence of different patterns is marked with different patches.  The blue line (coefficient of determination near unity) is the interface separating the disordered pattern (cyan, phase I) from the coherent pattern (white, phase II). The pattern is identified as coherent when the energetic mode becomes one of the last 10 nodes in the spectral energy diagram. The simulation is conducted with $N=512$, $\phi=0.01$ and $R_o=4.5$.}
\label{fig:map01} 
\end{figure}

Deterministic spatial patterns occur when energetic modes in the Fourier spectra are restricted to a narrow band of wavenumbers. They are stationary when the aforementioned narrow range of energetic modes do not vary in time.  Such energetic modes are determined here from the maxima of the squared Fourier amplitudes during the course of disease spread in dimensionless time ($\gamma t$). For an arbitrary kernel cut off of the 1-D system, as shown elsewhere \citep{fuentes2003nonlocal}, pattern separation occurs at the scale $\sigma/L=2/9$ for a 1-D system. A different pattern separation result is also observed in Figure \ref{fig:map01} for the 2-D SIR system with $R_o>2$ at $t_p$.  

As shown in Figure \ref{fig:map01}, kernel size can influence the type of spatial pattern setting at $t_p$. When $R_a/dr<5$, disordered spatial patterning always forms initially (Phase I) for a wide range of $R_a/\sigma$. For $R_a/\sigma>3$ values, the separation between the disordered (Phase I) and coherent (Phase II) patterns follows an approximate linear relation with $\sigma/dr \approx 5$.  Hence, the transition between disordered and coherent pattern at $t_p$ depends on the shape (i.e. $\sigma/dr$, normalized standard deviation) of the kernel for such kernel-based IDE \citep{fuentes2003nonlocal}. 

\section{Discussion}
\begin{figure} %[!ht] 
\centerline{\includegraphics [angle=0,height=0.54\linewidth]{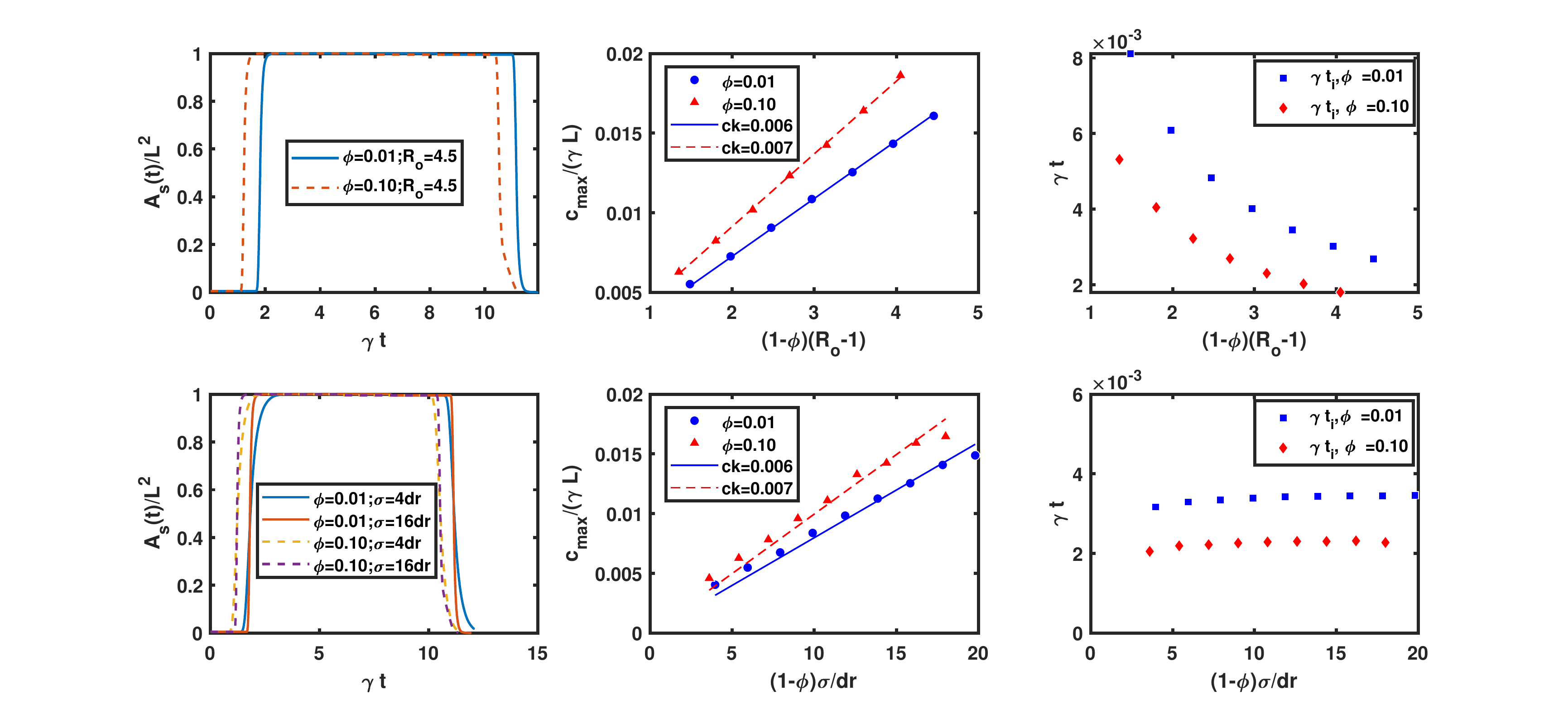}}
\caption{Left column: Variation of the spreading area $A_s$ in dimensionless time $\gamma t$ normalized by the domain size $L^2$.  Middle column: maximal expansion speed variations with $\phi$, $R_o$, and kernel spreading property $\sigma$. Right column: Normalized time to peak infections as a function of $\phi$, $R_o$, and $\sigma$. Randomized initial conditions are used.  The other input model parameters are identical to Figure \ref{fig:speed}.}
\label{fig:speedp} 
\end{figure}

In Figure \ref{fig:speed}, all infected individuals are placed at the center of the domain thus allowing a single wavefront to develop, expand, then contract at long times. Here, random initial conditions in space are used to explore whether the 1-D scaling analysis holds even when the expanding waves interfere. The expansion speeds are examined in Figure \ref{fig:speedp}.
Figure \ref{fig:speedp} shows that when random initial conditions are imposed spatially, the magnitudes of the expansion speeds increase with increasing $R_o$ and $\sigma$, as expected from the 1-D analysis. The linear relation still holds, which indicates that even when several traveling waves form and later interfere with each other, the overall expansion speed remains linear in $R_o$. The $c_{max}$ also occurs earlier when compared to a single infection initial condition at the center of the domain (as expected). Moreover,  $\phi$ has a small impact on the magnitude of the maximum velocity but acts as a retardation factor to the attainment of $c_{max}$ as before. Unsurprisingly, the time to peak infection also decreases accordingly with increasing $c_{max}$ and increasing $\phi$. The coefficient $c_k$ is higher than those featured in Figure \ref{fig:speed}, indicating dependency on initial conditions. 

\begin{figure} %[!ht] 
\centerline{\includegraphics [angle=0,height=0.63\linewidth]{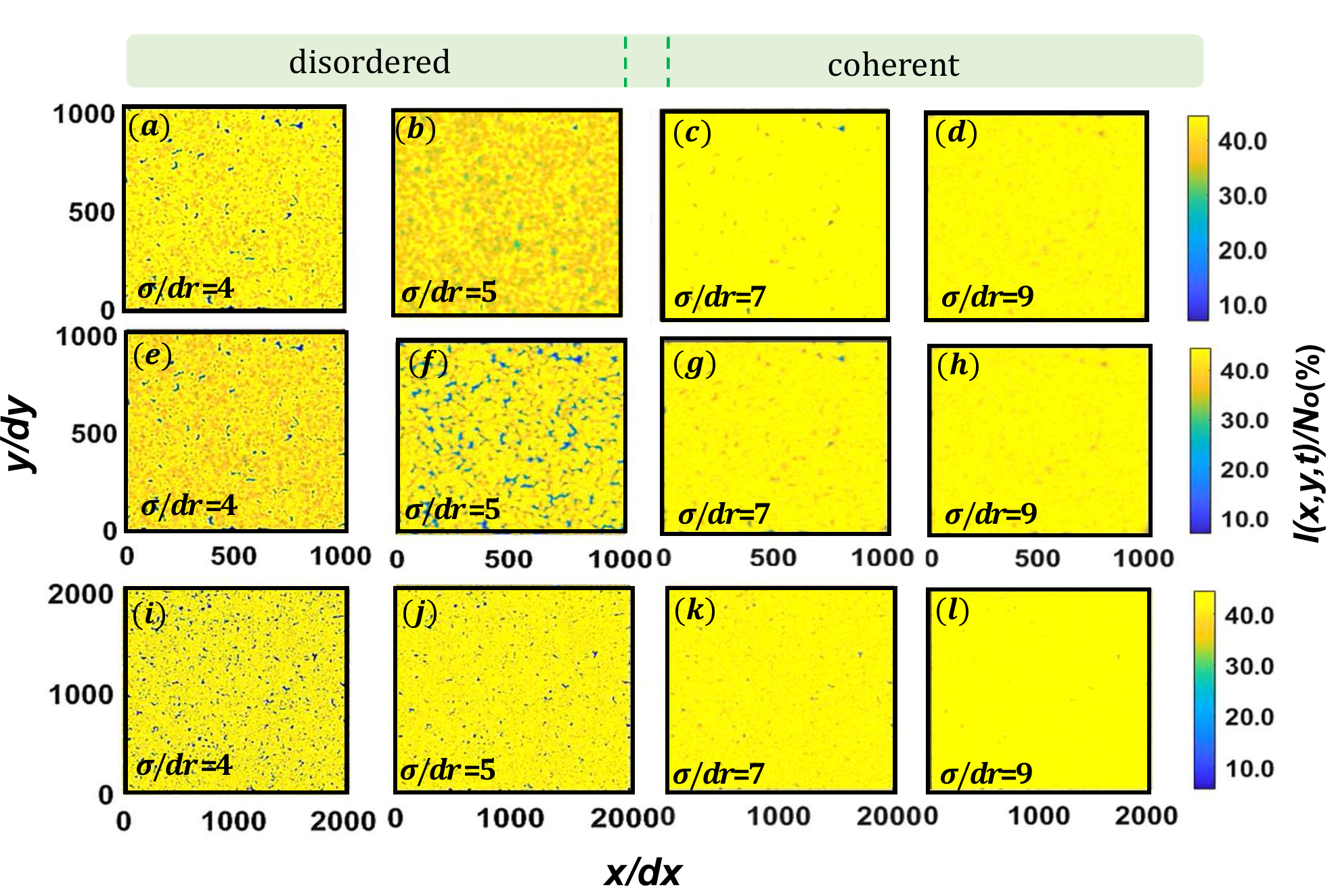}}
\caption{Evolution of patterns at $t_p$ when the domain size is enlarged. It is noted that the kernel sizes in the three rows are $R_a/dr=256$, $64$ and $256$, respectively. The patterns shown in the first two columns have clear disordered structures where as the patterns observed in the last two columns are coherent.}
\label{fig:p1024} 
\end{figure}

While $\phi$ is small (but finite), the developing infectious area occupies the entire domain.  Hence, infectious individuals acting through $\phi$ may be termed as geographical spreaders because only their mobility results in disease expansion that impacts the entire spread in the domain area. Last, variations in $\sigma/dr$ lead to a linear regime for intermediate values of $\sigma/dr>5$ again consistent with predictions from equation \ref{eq:SIR14}. Around $\sigma/dr\approx 5$, $c_{max}$ in both Figure \ref{fig:speed} and Figure \ref{fig:speedp} suggest a departure from linearity. This threshold $\sigma/dr$ is consistent with the pattern transition results in Figure \ref{fig:map01}.  Hence, while $\langle I\rangle $ dynamics are not sensitive to $\sigma/dr$, $c_{max}$ does sense the spatial pattern in SIR dynamics. To test the stability of this transition at $\sigma/dr \approx 5$, systems with different domain sizes are also investigated.  The simulation results with a domain size of $N=$ 1024 and 2048 are shown in Figure \ref{fig:p1024}. It is noted here the randomly initialized conditions and $R_o=4.5$ are the same as Figure \ref{fig:map01}.  Figure \ref{fig:p1024} shows that the transition between the disordered pattern and the coherent structure remains persistent at $\sigma/dr \approx 5$, regardless of the domain size enlargement from N=512 to 1024 and 2048.  

\section{Summary and Conclusion}

A kernel-based model is used to investigate the impact of geographical spreaders (i.e., highly mobile infected individuals) across identical communities - each governed by identical SIR dynamics. Hence, mass action between susceptible and infectious individuals is still assumed to occur locally within communities whereas mobility across communities is restricted to a small fraction of infectious individuals per unity time ($\phi$). This representation leads to a kernel-based SIR, where the infectious population is governed by an integro-differential equation (IDE).  The IDE allows for the description of disease spread in continuous instead of discrete time and in space when the size of the communities is much smaller than the domain size. Moreover, the proposed IDE accommodates non-local spatial processes when compared to diffusive transfer. The resulting IDE is shown to be unstable to small perturbations at any scale around the equilibrium state for early times, which is a necessary (but not sufficient) condition for pattern formation in space.  Scaling analysis (in 1-D) within an unbounded domain and numerical simulations (in 2-D) for a bounded domain with periodic boundary conditions indicate that the maximum spreading speed of the disease is determined by conventional epidemiological parameters $\gamma$ and $R_o$, and on the spatial mobility kernel properties of geographical spreaders, including its standard deviation $\sigma$ and finite-size $R_a$. The small $\phi$ does not alter the maximal spreading speed, but can retard the time needed to attain this maximum.  Numerical simulations confirm that when the standard deviation of the spreading kernel is larger than a critical value, spatial patterns can change from disordered to coherent structures at the time of peak infections in the domain.  This change resembles phase-transitions observed in other IDE systems.  Future work will consider (i) the role of the kernel shape (exponential, Laplace, and Wald) itself, (ii) the initial spatial distribution of the susceptible population within the lattice.  The latter is known to be reasonably represented by a multi-fractal process. 

The results reported here may be heuristic in guiding the early prevention of the epidemics that can be described by the SIR model. Reducing the kernel size, equivalent to constraining the movement of the geographical spreaders, is beneficial for reducing the maximal epidemic spreading speed and retarding the fast expansion of the epidemics. Similarly, controlling the variance of the spatial kernel can act to prevent the formation of coherent patterns (simultaneous breakouts) in the entire domain, which can significantly leverage the burden of the hospital system across communities.

\section*{Acknowledgments}
 G. Katul acknowledges support from the U.S. National Science Foundation (NSF-AGS-1644382, NSF-AGS-2028633, and NSF-IOS-1754893). S. Li acknowledges the financial support from the Nicholas School of the Environment at Duke University.

\section*{Appendix}
\renewcommand{\thefigure}{A\arabic{figure}}
\setcounter{figure}{0}
\renewcommand{\thetable}{A\arabic{table}}
\setcounter{table}{0}
\renewcommand{\theequation}{A\arabic{equation}}
\setcounter{equation}{0}

\begin{figure} %[ht]  
\centerline{\includegraphics [angle=0,height=0.33\linewidth]{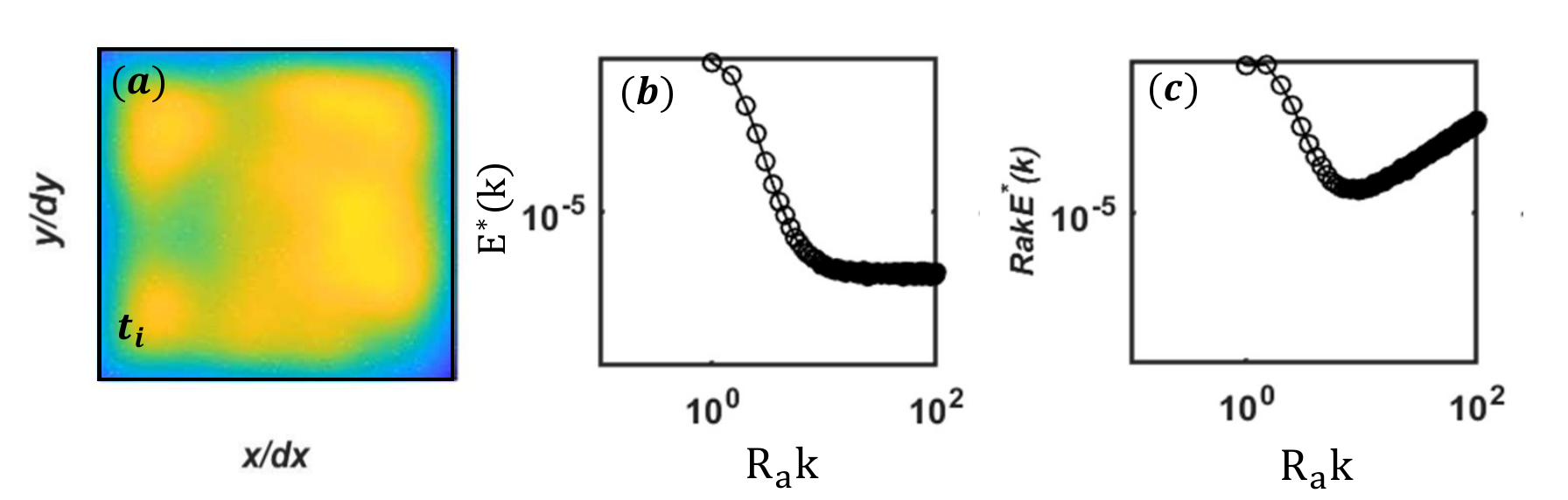}}
\caption{The spectral density $E^*(k)$ and cumulative energy spectra $R_a k E^*(k)$ of the pseudo-diffusive pattern at $t_i$. The simulation parameters are identical to plot (i) of Figure \ref{fig:sirt}.}
\label{fig:sirt2} 
\end{figure}

\begin{figure} %[ht] 
\centerline{\includegraphics [angle=0,height=0.33\linewidth]{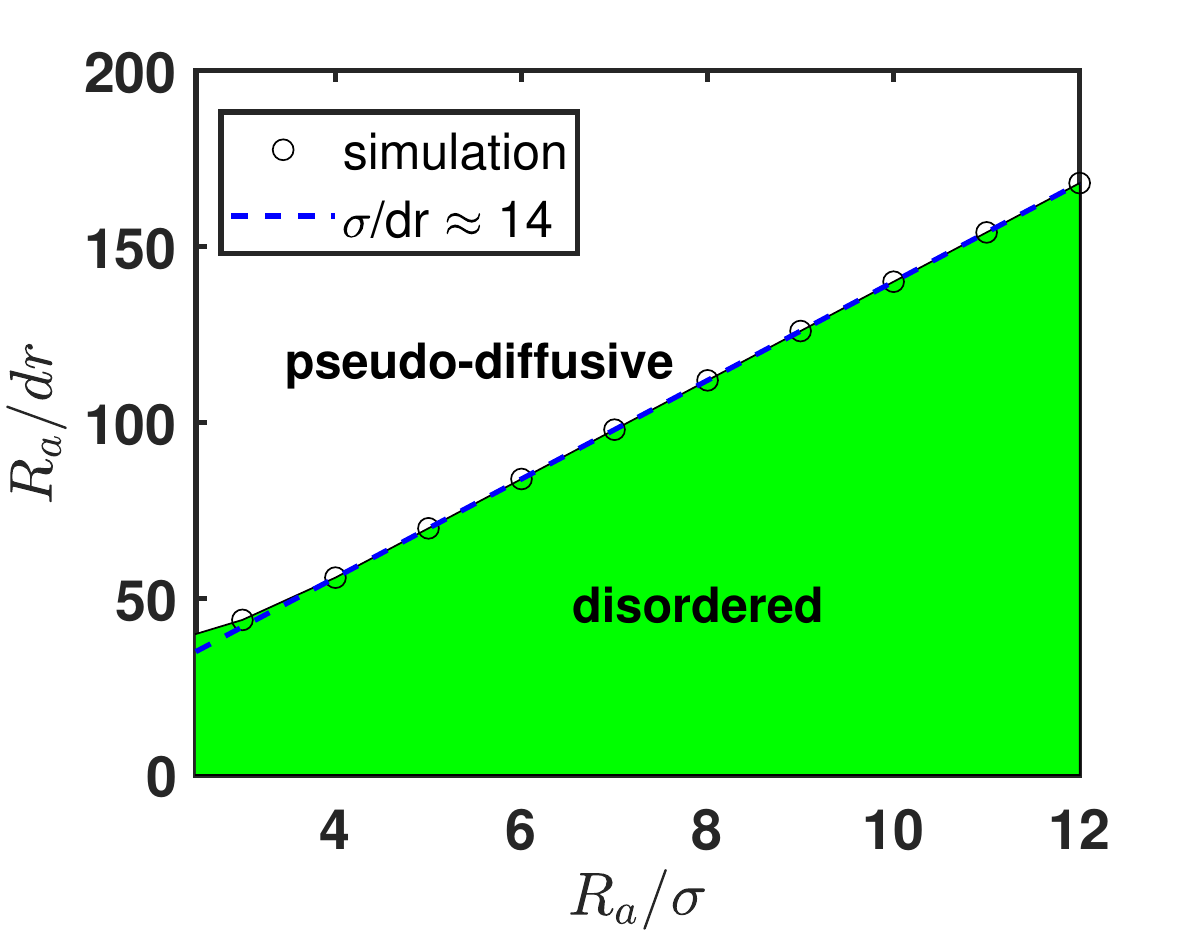}}
\caption{Spatial map for the pattern transition at time $t_i$. The occurrence of different patterns is marked with with different patches. The blue line (coefficient of determination approximately unity) is the interface separating the disordered pattern (green) from the pseudo-diffusive pattern (white). The simulation is conducted with $N=512$, $\phi=0.01$ and $R_o=4.5$.}
\label{fig:map02} 
\end{figure}

As shown in Figure \ref{fig:sirt}, the pseudo-diffusive pattern appears at $t_i$ when $d\langle I\rangle /dt$ is maximum. This pattern is (i) transient since the initial condition still dominates the entropy plot and (ii) sensitive to the domain size. The appendix here shows that the transition from disordered to pseudo-diffusive is also governed by a phase-transition depending on the kernel properties. The spectral features of this pattern are shown in Figure \ref{fig:sirt2}. The pattern partition map between disordered and pseudo-diffusive is now shown in Figure \ref{fig:map02}.

\begin{figure} %[!ht] 
\centerline{\includegraphics [angle=0,height=0.63\linewidth]{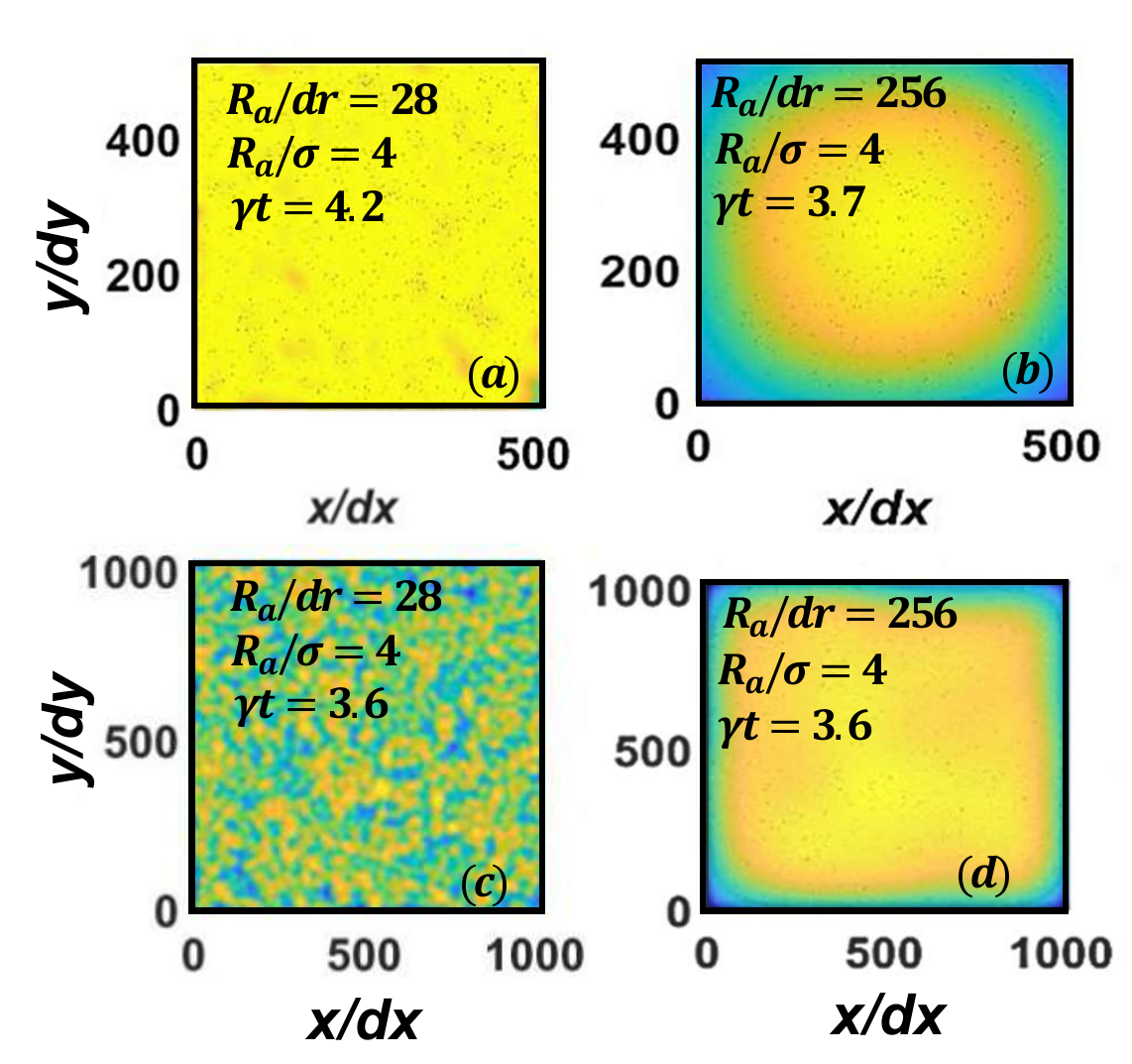}}
\caption{Comparison of the pseudo-diffusive patterns when the domain size doubles. Plots (a) and (b) show two different pseudo-diffusive patterns at $t_i$ when $N=512$. The two patterns in plots (c) and (d) correspond to a disordered (panel c) and pseudo-diffusive (panel d) when the domain size doubles (N=1024).}
\label{fig:Dp1024} 
\end{figure}

The spectral structure of the pseudo-diffusive pattern shown in Figure \ref{fig:sirt2} is similar to the coherent structure in Figure \ref{fig:sirt}. However, the energetic mode of the pseudo-diffusive pattern of $R_a k E^*(k)$ emerges in the first few wavenumbers. This suggests that the pattern is dictated by the largest length scale in the domain. To test whether its morphology (i.e. labyrinth-like shape) is influenced by the domain size, further simulations with $N=1024$ are conducted and shown in Figure \ref{fig:Dp1024}.

It is shown in Figure \ref{fig:Dp1024} that when the domain size changes, the maximal energetic modes at $t_i$ also changes, confirming that the pseudo-diffusive pattern is sensitive to the domain size. This is consistent with the spectral features observed in Figure \ref{fig:sirt2}. However, the more pertinent finding is that for both $N=512$ and $N=1024$, the pattern morphology (i.e. pseudo-diffusive classification) still holds despite domain size doubling.

\bibliography{2_SIR_BIB}
\end{document}